\documentclass{article}

\usepackage{arxiv}

\usepackage[utf8]{inputenc} 
\usepackage[T1]{fontenc}    
\usepackage{hyperref}       
\usepackage{url}            
\usepackage{booktabs}       
\usepackage{amsfonts}       
\usepackage{nicefrac}       
\usepackage{microtype}      
\usepackage{lipsum}
\usepackage{graphicx}

\title{snpQT: flexible, reproducible, and comprehensive quality control and imputation of genomic data}

\author{
  Christina Vasilopoulou\\
  Northern Ireland Centre for Stratified Medicine\\
  Altnagelvin Hospital Campus\\
  Biomedical Sciences Research Institute \\
  Ulster University\\
  Derry/Londonderry, UK, BT47 6SB\\
  \texttt{vasilopoulou-c@ulster.ac.uk} \\
  \And
  Benjamin Wingfield \\
  Centre for Personalised Medicine\\
  Altnagelvin Hospital Campus\\
  Biomedical Sciences Research Institute\\
  Ulster University\\
  Derry/Londonderry, UK, BT47 6SB \\
  \texttt{b.wingfield@ulster.ac.uk} \\
  \AND
  Andrew P. Morris \\
  Centre for Genetics and Genomics Versus Arthritis\\
  Centre for Musculoskeletal Research\\
  Manchester Academic Health Science Centre\\
  University of Manchester\\
  Manchester, M13 9PT, UK\\
  \texttt{andrew.morris-5@manchester.ac.uk} \\
  \And
  William Duddy \\
  Northern Ireland Centre for Stratified Medicine\\
  Altnagelvin Hospital Campus\\
  Biomedical Sciences Research Institute \\
  Ulster University\\
  Derry/Londonderry, UK, BT47 6SB\\
  \texttt{w.duddy@ulster.ac.uk} \\
}

\begin{document}
\maketitle

\begin{abstract}
\textbf{Motivation:} Quality control of genomic data is an essential but complicated multi-step procedure, often requiring separate installation and expert familiarity with a combination of disparate bioinformatics tools.\\
\textbf{Results:} To provide an automated solution that retains comprehensive quality checks and flexible workflow architecture, we have developed snpQT, a scalable, stand-alone software pipeline, offering some 36 discrete quality filters or correction steps, with  plots before-and-after user-modifiable thresholding. This includes build conversion, population stratification against 1,000 Genomes data, population outlier removal, and built-in imputation with its own pre- and post- quality controls. Common input formats are used and users need not be superusers nor have any prior coding experience. A comprehensive online tutorial and installation guide is provided through to GWAS (\texttt{https://snpqt.readthedocs.io/en/latest/}), introducing snpQT using a synthetic demonstration dataset and a real-world Amyotrophic Lateral Sclerosis SNP-array dataset.\\
\textbf{Availability:} snpQT is open source and freely available at \texttt{https://github.com/nebfield/snpQT}. \\
\textbf{Contact:} \href{Vasilopoulou-C@ulster.ac.uk}{Vasilopoulou-C@ulster.ac.uk}, \href{w.duddy@ulster.ac.uk}{w.duddy@ulster.ac.uk}\\
\end{abstract}


\section{Introduction}
 Assuring high quality of genomic data is necessarily a complex multi-step procedure, but it is critical to generate reproducible and reliable results in genome-wide association studies (GWAS). Multiple challenges are encountered in carrying out QC  (\cite{Vasilopoulou2021ALSML}). Although there are well-established steps and good practices (\cite{Anderson2011,Marees2018}), there is no standardised and universally followed workflow, contributing to low reproducibility of results.

Existing approaches, including semi-automated tools (\cite{odyssey19}), can involve a time-consuming "trial and error" approach, requiring the analyst to check the distributions of parameters in plots produced over many rounds of adjustments, and to manually enter commands in a long list of QC steps one-by-one or in a series of shell scripts. The analyst may encounter incompatibility problems and installation difficulties. Software architecture tools such as nextflow and BioContainers can address these issues and have been proposed as automated solutions (\cite{nf-gwas21}), but limitations exist in terms of limited and relatively rigid QC analysis, lacking such steps as imputation, limited variety of threshold choice and plot outputs, and the requirement for users to have extensive knowledge of the software in order to tailor their analysis.

\begin{figure}
\centering
\includegraphics[width=17 cm]{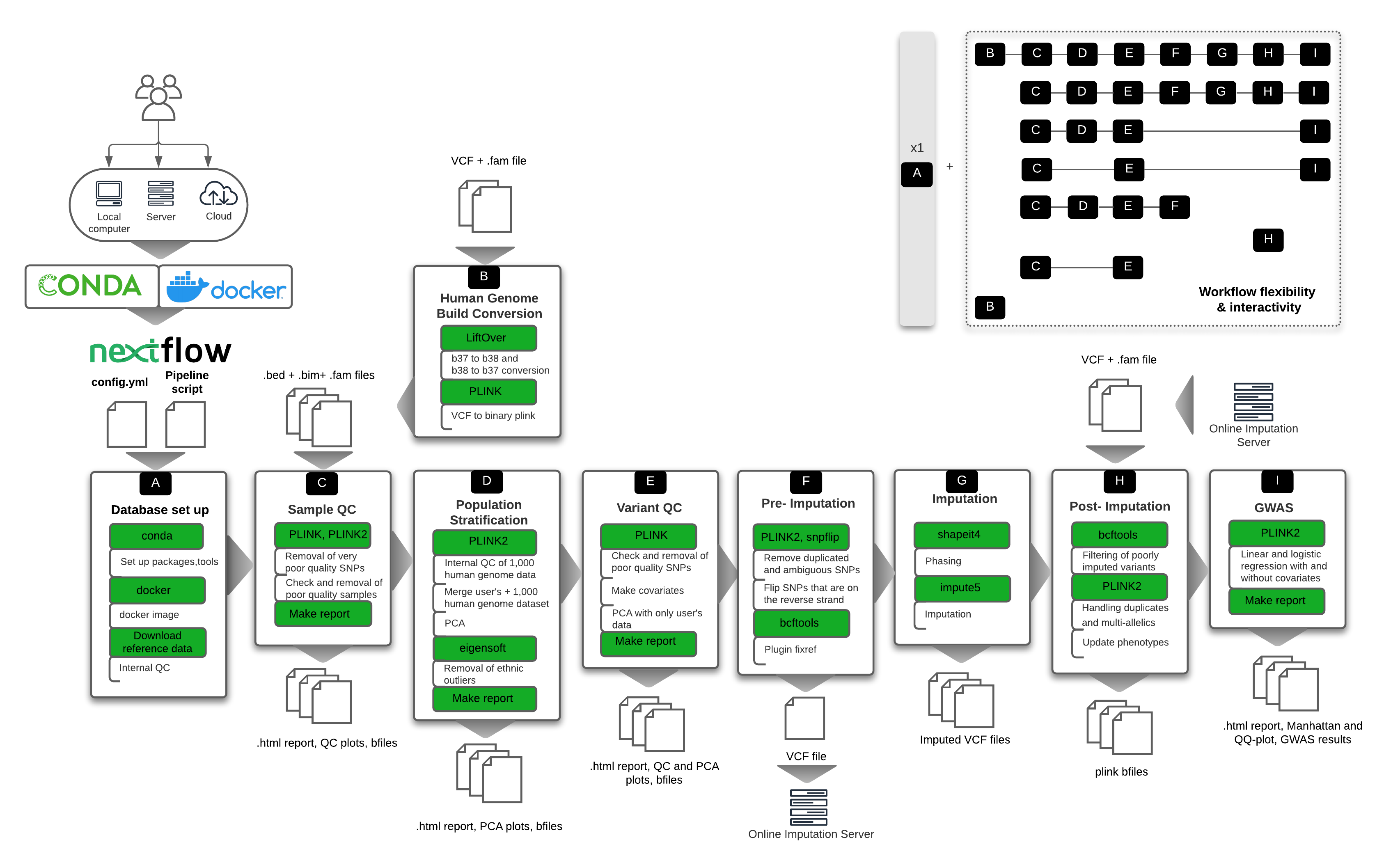}
\caption{Outline of the snpQT architecture, which includes nine core workflows (A-I) that are implemented using nextflow. Each workflow expects specific inputs either from the user or from the outputs generated by other workflows. Main tools and key processes (modules) are highlighted in green.  Examples of different task combinations are represented in the upper-right corner, showing the flexibility and interactivity among the implemented workflows. VCF: Variant Call File; QC: Quality Control; PCA: Principal Component Analysis, bfiles: Binary PLINK files.}
\label{fig:fig1}
\end{figure}

\section{Methods}
\label{sec:headings}
snpQT was developed as a set of nine core workflow components implemented with the nextflow workflow management system (\cite{di2017nextflow}). Each workflow component consists of independent containerised modules, using BioContainers curated by the bioinformatics community wherever possible (\cite{da2017biocontainers}). Nextflow allows snpQT to be easily scaled from a laptop to a high-performance computing (HPC) or cloud environment, and enables caching at continuous checkpoints, so users can alter thresholds without needing to rerun earlier parts of the analysis.

All nine workflows are illustrated in Figure~1\vphantom{\ref{fig:01}}. Workflow A runs only once, performing a local database set up, downloading and preparing reference files (\cite{1000genomes,plink21k}) and setting up specific versions of tools using conda or docker. snpQT processes data in human genome build 37, but Workflow B has been created for the user to convert from build 38 to 37 or vice versa. Workflow C performs sample QC, including checks for missing call rate, sex discrepancies, heterozygosity, cryptic relatedness, and missing phenotypes. Workflow D performs population stratification for the automatic removal of samples that are predicted as ethnic outliers (using EIGENSOFT, \cite{eigensoft06}). Workflow E performs the main Variant QC, checking missing call rate, Hardy-Weinberg equilibrium deviation, minor allele frequency, missingness in case/control status, and generates covariates for GWAS, based on a user-modifiable number of Principal Components (or users may provide a covariates file). Workflow F is for pre-imputation quality control, while workflow G performs local phasing and imputation using shapeit4 (\cite{shapeit4}) and impute5 (\cite{impute5}), and workflow H performs post-imputation QC. The workflows structure also allows for users to upload their data to an external imputation server, or use a different reference panel.  Workflow I performs GWAS, outputting summary statistics, along with a Manhattan plot and a QQ-plot. Detailed summary logs and graphs are provided throughput, depicting the total number of samples and variants in each step, and prompting users towards the locations of intermediate files and logs.

snpQT is implemented in nextflow, R and bash. As well as those already listed, the following tools are used: picard (\texttt{https://broadinstitute.github.io/picard/}), PLINK (\cite{plink}), PLINK2.0 (\cite{plink2}), samtools (\cite{samtools21}), and snpflip (\texttt{https://github.com/biocore-ntnu/snpflip}). The latest release of 1,000 human genome data (\cite{1000genomes}) is used as a reference panel in both VCF and processed PLINK2 formats (\cite{plink21k}). A part of population stratification implementation was inspired by the work of \cite{Marees2018}.

We demonstrate snpQT using a synthetic dataset which is available with the tool, and an Amyotrophic Lateral Sclerosis SNP-array dataset of 2,000 samples (1,000 cases and 1,000 controls) taken from a restricted-access dbGaP project (\cite{nicolas2018}), at \texttt{https://snpqt.readthedocs.io/en/latest/}.  

\section{Conclusion}

snpQT offers robust QC combined with scalability, reproducibility, flexibility and user-friendly design which can appeal to a broad spectrum of users. It is stand-alone software that needs neither additional coding nor manual installation/download of any data or other program apart from nextflow and conda or docker. The input is a VCF file and/or binary plink files, formats which are widely used. For users who have limited experience with QC analysis a thorough "how-to" guide and step-by-step tutorials are provided, using the demonstration dataset that is available with the tool.

\section*{Acknowledgements}

We would like to thank Dr. Priyank Shukla for helpful discussion, Dr. Apostolos Malatras for his assistance, and Peter Timlett for designing the snpQT logo.

\section*{Funding}

This work was financed by the EU Regional Development Fund EU Sustainable Competitiveness Programme for N. Ireland, NI Public Health Agency (HSC R\&D) \& Ulster University. C.V. was the recipient of a DfE international scholarship from Ulster University.

\bibliographystyle{unsrt}  

\bibliography{references}

\begin{thebibliography}{10}

\bibitem{Vasilopoulou2021ALSML}
Christina Vasilopoulou, Andrew~P. Morris, George Giannakopoulos, Stephanie
  Duguez, and William Duddy.
\newblock {What Can Machine Learning Approaches in Genomics Tell Us about the
  Molecular Basis of Amyotrophic Lateral Sclerosis?}
\newblock {\em Journal of Personalized Medicine}, 10(4):247, 11 2020.

\bibitem{Anderson2011}
Carl~A Anderson, Fredrik~H Pettersson, Geraldine~M Clarke, Lon~R Cardon,
  P~Morris, and Krina~T Zondervan.
\newblock {Data quality control in genetic case-control association studies}.
\newblock {\em Nature Protocols}, 5(9):1564--1573, 2011.

\bibitem{Marees2018}
Andries~T. Marees, Hilde de~Kluiver, Sven Stringer, Florence Vorspan, Emmanuel
  Curis, Cynthia Marie-Claire, and Eske~M. Derks.
\newblock {A tutorial on conducting genome-wide association studies: Quality
  control and statistical analysis}.
\newblock {\em International Journal of Methods in Psychiatric Research},
  27(2):1--10, 2018.

\bibitem{odyssey19}
Ryan~J. Eller, Sarath~C. Janga, and Susan Walsh.
\newblock {Odyssey: A semi-automated pipeline for phasing, imputation, and
  analysis of genome-wide genetic data}.
\newblock {\em BMC Bioinformatics}, 20(1):364, 6 2019.

\bibitem{nf-gwas21}
Zeyuan Song, Anastasia Gurinovich, Anthony Federico, Stefano Monti, and Paola
  Sebastiani.
\newblock {nf-gwas-pipeline: A Nextflow Genome-Wide Association Study
  Pipeline}.
\newblock {\em Journal of Open Source Software}, 6(59):2957, 3 2021.

\bibitem{di2017nextflow}
Paolo DI~Tommaso, Maria Chatzou, Evan~W. Floden, Pablo~Prieto Barja, Emilio
  Palumbo, and Cedric Notredame.
\newblock {Nextflow enables reproducible computational workflows}, 4 2017.

\bibitem{da2017biocontainers}
Felipe da~Veiga~Leprevost, Björn~A. Gr{\"{u}}ning, Saulo Alves~Aflitos,
  Hannes~L. R{\"{o}}st, Julian Uszkoreit, Harald Barsnes, Marc Vaudel, Pablo
  Moreno, Laurent Gatto, Jonas Weber, Mingze Bai, Rafael~C. Jimenez, Timo
  Sachsenberg, Julianus Pfeuffer, Roberto Vera~Alvarez, Johannes Griss,
  Alexey~I. Nesvizhskii, and Yasset Perez-Riverol.
\newblock {BioContainers: an open-source and community-driven framework for
  software standardization}.
\newblock {\em Bioinformatics (Oxford, England)}, 33(16):2580--2582, 8 2017.

\bibitem{1000genomes}
Adam Auton, Gonçalo~R. Abecasis, David~M. Altshuler, Richard~M. Durbin,
  David~R. Bentley, Aravinda Chakravarti, Andrew~G. Clark, Peter Donnelly,
  Evan~E. Eichler, Paul Flicek, Stacey~B. Gabriel, Richard~A. Gibbs, Eric~D.
  Green, Matthew~E. Hurles, Bartha~M. Knoppers, Jan~O. Korbel, Eric~S. Lander,
  Charles Lee, Hans Lehrach, Elaine~R. Mardis, Gabor~T. Marth, Gil~A. McVean,
  Deborah~A. Nickerson, Jeanette~P. Schmidt, Stephen~T. Sherry, Jun Wang,
  Richard~K. Wilson, Eric Boerwinkle, Harsha Doddapaneni, Yi~Han, Viktoriya
  Korchina, Christie Kovar, Sandra Lee, Donna Muzny, Jeffrey~G. Reid, Yiming
  Zhu, Yuqi Chang, Qiang Feng, Xiaodong Fang, Xiaosen Guo, Min Jian, Hui Jiang,
  Xin Jin, Tianming Lan, Guoqing Li, Jingxiang Li, Yingrui Li, Shengmao Liu,
  Xiao Liu, Yao Lu, Xuedi Ma, Meifang Tang, Bo~Wang, Guangbiao Wang, Honglong
  Wu, Renhua Wu, Xun Xu, Ye~Yin, Dandan Zhang, Wenwei Zhang, Jiao Zhao, Meiru
  Zhao, Xiaole Zheng, Namrata Gupta, Neda Gharani, Lorraine~H. Toji, Norman~P.
  Gerry, Alissa~M. Resch, Jonathan Barker, Laura Clarke, Laurent Gil, Sarah~E.
  Hunt, Gavin Kelman, Eugene Kulesha, Rasko Leinonen, William~M. McLaren,
  Rajesh Radhakrishnan, Asier Roa, Dmitriy Smirnov, Richard~E. Smith, Ian
  Streeter, Anja Thormann, Iliana Toneva, Brendan Vaughan, Xiangqun
  Zheng-Bradley, Russell Grocock, Sean Humphray, Terena James, Zoya Kingsbury,
  Ralf Sudbrak, Marcus~W. Albrecht, Vyacheslav~S. Amstislavskiy, Tatiana~A.
  Borodina, Matthias Lienhard, Florian Mertes, Marc Sultan, Bernd Timmermann,
  Marie~Laure Yaspo, Lucinda Fulton, Victor Ananiev, Zinaida Belaia, Dimitriy
  Beloslyudtsev, Nathan Bouk, Chao Chen, Deanna Church, Robert Cohen, Charles
  Cook, John Garner, Timothy Hefferon, Mikhail Kimelman, Chunlei Liu, John
  Lopez, Peter Meric, Chris O'Sullivan, Yuri Ostapchuk, Lon Phan, Sergiy
  Ponomarov, Valerie Schneider, Eugene Shekhtman, Karl Sirotkin, Douglas
  Slotta, Hua Zhang, Senduran Balasubramaniam, John Burton, Petr Danecek,
  Thomas~M. Keane, Anja Kolb-Kokocinski, Shane McCarthy, James Stalker, Michael
  Quail, Christopher~J. Davies, Jeremy Gollub, Teresa Webster, Brant Wong,
  Yiping Zhan, Christopher~L. Campbell, Yu~Kong, Anthony Marcketta, Fuli Yu,
  Lilian Antunes, Matthew Bainbridge, Aniko Sabo, Zhuoyi Huang, Lachlan~J.M.
  Coin, Lin Fang, Qibin Li, Zhenyu Li, Haoxiang Lin, Binghang Liu, Ruibang Luo,
  Haojing Shao, Yinlong Xie, Chen Ye, Chang Yu, Fan Zhang, Hancheng Zheng,
  Hongmei Zhu, Can Alkan, Elif Dal, Fatma Kahveci, Erik~P. Garrison, Deniz
  Kural, Wan~Ping Lee, Wen~Fung Leong, Michael Stromberg, Alistair~N. Ward,
  Jiantao Wu, Mengyao Zhang, Mark~J. Daly, Mark~A. DePristo, Robert~E.
  Handsaker, Eric Banks, Gaurav Bhatia, Guillermo Del~Angel, Giulio Genovese,
  Heng Li, Seva Kashin, Steven~A. McCarroll, James~C. Nemesh, Ryan~E. Poplin,
  Seungtai~C. Yoon, Jayon Lihm, Vladimir Makarov, Srikanth Gottipati, Alon
  Keinan, Juan~L. Rodriguez-Flores, Tobias Rausch, Markus~H. Fritz, Adrian~M.
  St{\"{u}}tz, Kathryn Beal, Avik Datta, Javier Herrero, Graham~R.S. Ritchie,
  Daniel Zerbino, Pardis~C. Sabeti, Ilya Shlyakhter, Stephen~F. Schaffner,
  Joseph Vitti, David~N. Cooper, Edward~V. Ball, Peter~D. Stenson, Bret Barnes,
  Markus Bauer, R.~Keira Cheetham, Anthony Cox, Michael Eberle, Scott Kahn,
  Lisa Murray, John Peden, Richard Shaw, Eimear~E. Kenny, Mark~A. Batzer,
  Miriam~K. Konkel, Jerilyn~A. Walker, Daniel~G. MacArthur, Monkol Lek, Ralf
  Herwig, Li~Ding, Daniel~C. Koboldt, David Larson, Kai Ye, Simon Gravel, Anand
  Swaroop, Emily Chew, Tuuli Lappalainen, Yaniv Erlich, Melissa Gymrek,
  Thomas~Frederick Willems, Jared~T. Simpson, Mark~D. Shriver, Jeffrey~A.
  Rosenfeld, Carlos~D. Bustamante, Stephen~B. Montgomery, Francisco~M.
  De~La~Vega, Jake~K. Byrnes, Andrew~W. Carroll, Marianne~K. DeGorter, Phil
  Lacroute, Brian~K. Maples, Alicia~R. Martin, Andres Moreno-Estrada, Suyash~S.
  Shringarpure, Fouad Zakharia, Eran Halperin, Yael Baran, Eliza Cerveira,
  Jaeho Hwang, Ankit Malhotra, Dariusz Plewczynski, Kamen Radew, Mallory
  Romanovitch, Chengsheng Zhang, Fiona~C.L. Hyland, David~W. Craig, Alexis
  Christoforides, Nils Homer, Tyler Izatt, Ahmet~A. Kurdoglu, Shripad~A.
  Sinari, Kevin Squire, Chunlin Xiao, Jonathan Sebat, Danny Antaki, Madhusudan
  Gujral, Amina Noor, Kenny Ye, Esteban~G. Burchard, Ryan~D. Hernandez,
  Christopher~R. Gignoux, David Haussler, Sol~J. Katzman, W.~James Kent, Bryan
  Howie, Andres Ruiz-Linares, Emmanouil~T. Dermitzakis, Scott~E. Devine,
  Hyun~Min Kang, Jeffrey~M. Kidd, Tom Blackwell, Sean Caron, Wei Chen, Sarah
  Emery, Lars Fritsche, Christian Fuchsberger, Goo Jun, Bingshan Li, Robert
  Lyons, Chris Scheller, Carlo Sidore, Shiya Song, Elzbieta Sliwerska, Daniel
  Taliun, Adrian Tan, Ryan Welch, Mary~Kate Wing, Xiaowei Zhan, Philip
  Awadalla, Alan Hodgkinson, Yun Li, Xinghua Shi, Andrew Quitadamo, Gerton
  Lunter, Jonathan~L. Marchini, Simon Myers, Claire Churchhouse, Olivier
  Delaneau, Anjali Gupta-Hinch, Warren Kretzschmar, Zamin Iqbal, Iain
  Mathieson, Androniki Menelaou, Andy Rimmer, Dionysia~K. Xifara, Taras~K.
  Oleksyk, Yunxin Fu, Xiaoming Liu, Momiao Xiong, Lynn Jorde, David
  Witherspoon, Jinchuan Xing, Brian~L. Browning, Sharon~R. Browning, Fereydoun
  Hormozdiari, Peter~H. Sudmant, Ekta Khurana, Chris Tyler-Smith, Cornelis~A.
  Albers, Qasim Ayub, Yuan Chen, Vincenza Colonna, Luke Jostins, Klaudia
  Walter, Yali Xue, Mark~B. Gerstein, Alexej Abyzov, Suganthi Balasubramanian,
  Jieming Chen, Declan Clarke, Yao Fu, Arif~O. Harmanci, Mike Jin, Donghoon
  Lee, Jeremy Liu, Xinmeng~Jasmine Mu, Jing Zhang, Yan Zhang, Chris Hartl,
  Khalid Shakir, Jeremiah Degenhardt, Sascha Meiers, Benjamin Raeder,
  Francesco~Paolo Casale, Oliver Stegle, Eric~Wubbo Lameijer, Ira Hall, Vineet
  Bafna, Jacob Michaelson, Eugene~J. Gardner, Ryan~E. Mills, Gargi Dayama, Ken
  Chen, Xian Fan, Zechen Chong, Tenghui Chen, Mark~J. Chaisson, John
  Huddleston, Maika Malig, Bradley~J. Nelson, Nicholas~F. Parrish, Ben
  Blackburne, Sarah~J. Lindsay, Zemin Ning, Yujun Zhang, Hugo Lam, Cristina
  Sisu, Danny Challis, Uday~S. Evani, James Lu, Uma Nagaswamy, Jin Yu, Wangshen
  Li, Lukas Habegger, Haiyuan Yu, Fiona Cunningham, Ian Dunham, Kasper Lage,
  Jakob~Berg Jespersen, Heiko Horn, Donghoon Kim, Rob Desalle, Apurva
  Narechania, Melissa~A.Wilson Sayres, Fernando~L. Mendez, G.~David Poznik,
  Peter~A. Underhill, David Mittelman, Ruby Banerjee, Maria Cerezo, Thomas~W.
  Fitzgerald, Sandra Louzada, Andrea Massaia, Fengtang Yang, Divya Kalra,
  Walker Hale, Xu~Dan, Kathleen~C. Barnes, Christine Beiswanger, Hongyu Cai,
  Hongzhi Cao, Brenna Henn, Danielle Jones, Jane~S. Kaye, Alastair Kent,
  Angeliki Kerasidou, Rasika Mathias, Pilar~N. Ossorio, Michael Parker,
  Charles~N. Rotimi, Charmaine~D. Royal, Karla Sandoval, Yeyang Su, Zhongming
  Tian, Sarah Tishkoff, Marc Via, Yuhong Wang, Huanming Yang, Ling Yang,
  Jiayong Zhu, Walter Bodmer, Gabriel Bedoya, Zhiming Cai, Yang Gao, Jiayou
  Chu, Leena Peltonen, Andres Garcia-Montero, Alberto Orfao, Julie Dutil,
  Juan~C. Martinez-Cruzado, Rasika~A. Mathias, Anselm Hennis, Harold Watson,
  Colin McKenzie, Firdausi Qadri, Regina LaRocque, Xiaoyan Deng, Danny Asogun,
  Onikepe Folarin, Christian Happi, Omonwunmi Omoniwa, Matt Stremlau, Ridhi
  Tariyal, Muminatou Jallow, Fatoumatta~Sisay Joof, Tumani Corrah, Kirk
  Rockett, Dominic Kwiatkowski, Jaspal Kooner, Tran~Tinh Hien, Sarah~J.
  Dunstan, Nguyen ThuyHang, Richard Fonnie, Robert Garry, Lansana Kanneh, Lina
  Moses, John Schieffelin, Donald~S. Grant, Carla Gallo, Giovanni Poletti,
  Danish Saleheen, Asif Rasheed, Lisa~D. Brooks, Adam~L. Felsenfeld, Jean~E.
  McEwen, Yekaterina Vaydylevich, Audrey Duncanson, Michael Dunn, and
  Jeffery~A. Schloss.
\newblock {A global reference for human genetic variation}, 9 2015.

\bibitem{plink21k}
{Chang CC}.
\newblock {1000 Genomes phase 3, phased and annotated data for use in plink2.0
  worked examples}.
\newblock {\em GigaScience Database}, 2018.

\bibitem{eigensoft06}
Alkes~L. Price, Nick~J. Patterson, Robert~M. Plenge, Michael~E. Weinblatt,
  Nancy~A. Shadick, and David Reich.
\newblock {Principal components analysis corrects for stratification in
  genome-wide association studies}.
\newblock {\em Nature Genetics}, 38(8):904--909, 8 2006.

\bibitem{shapeit4}
Olivier Delaneau, Jean~François Zagury, Matthew~R. Robinson, Jonathan~L.
  Marchini, and Emmanouil~T. Dermitzakis.
\newblock {Accurate, scalable and integrative haplotype estimation}.
\newblock {\em Nature Communications}, 10(1):1--10, 12 2019.

\bibitem{impute5}
Simone Rubinacci, Olivier Delaneau, and Jonathan Marchini.
\newblock {Genotype imputation using the Positional Burrows Wheeler Transform}.
\newblock {\em PLOS Genetics}, 16(11):e1009049, 11 2020.

\bibitem{plink}
Shaun Purcell, Benjamin Neale, Kathe Todd-Brown, Lori Thomas, Manuel~A.R.
  Ferreira, David Bender, Julian Maller, Pamela Sklar, Paul~I.W. De~Bakker,
  Mark~J. Daly, and Pak~C. Sham.
\newblock {PLINK: A tool set for whole-genome association and population-based
  linkage analyses}.
\newblock {\em American Journal of Human Genetics}, 81(3):559--575, 2007.

\bibitem{plink2}
Christopher~C Chang, Carson~C Chow, Laurent~CAM Tellier, Shashaank Vattikuti,
  Shaun~M Purcell, and James~J Lee.
\newblock {Second-generation PLINK: rising to the challenge of larger and
  richer datasets}.
\newblock {\em GigaScience}, 4(1):7, 12 2015.

\bibitem{samtools21}
Petr Danecek, James~K Bonfield, Jennifer Liddle, John Marshall, Valeriu Ohan,
  Martin~O Pollard, Andrew Whitwham, Thomas Keane, Shane~A McCarthy, Robert~M
  Davies, and Heng Li.
\newblock {Twelve years of SAMtools and BCFtools.}
\newblock {\em GigaScience}, 10(2):1--4, 2 2021.

\bibitem{nicolas2018}
Aude Nicolas, Kevin Kenna, Alan~E. Renton, Nicola Ticozzi, Faraz Faghri, Ruth
  Chia, Janice~A. Dominov, Brendan~J. Kenna, Mike~A. Nalls, Pamela Keagle,
  Alberto~M. Rivera, Wouter van Rheenen, Natalie~A. Murphy, Joke~J.F.A. van
  Vugt, Joshua~T. Geiger, Rick van~der Spek, Hannah~A. Pliner,
  {Shankaracharya}, Bradley~N. Smith, Giuseppe Marangi, Simon~D. Topp,
  Yevgeniya Abramzon, Athina~Soragia Gkazi, John~D. Eicher, Aoife Kenna,
  Francesco~O. Logullo, Isabella Simone, Giancarlo Logroscino, Fabrizio Salvi,
  Ilaria Bartolomei, Giuseppe Borghero, Maria~Rita Murru, Emanuela Costantino,
  Carla Pani, Roberta Puddu, Carla Caredda, Valeria Piras, Stefania Tranquilli,
  Stefania Cuccu, Daniela Corongiu, Maurizio Melis, Antonio Milia, Francesco
  Marrosu, Maria~Giovanna Marrosu, Gianluca Floris, Antonino Cannas, Margherita
  Capasso, Claudia Caponnetto, Gianluigi Mancardi, Paola Origone, Paola
  Mandich, Francesca~L. Conforti, Sebastiano Cavallaro, Gabriele Mora, Kalliopi
  Marinou, Riccardo Sideri, Silvana Penco, Lorena Mosca, Christian Lunetta,
  Giuseppe~Lauria Pinter, Massimo Corbo, Nilo Riva, Paola Carrera, Paolo
  Volanti, Jessica Mandrioli, Nicola Fini, Antonio Fasano, Lucio Tremolizzo,
  Alessandro Arosio, Carlo Ferrarese, Francesca Trojsi, Gioacchino Tedeschi,
  Maria~Rosaria Monsurr{\`{o}}, Giovanni Piccirillo, Cinzia Femiano, Anna
  Ticca, Enzo Ortu, Vincenzo La~Bella, Rossella Spataro, Tiziana Colletti,
  Mario Sabatelli, Marcella Zollino, Amelia Conte, Marco Luigetti, Serena
  Lattante, Marialuisa Santarelli, Antonio Petrucci, Maura Pugliatti, Angelo
  Pirisi, Leslie~D. Parish, Patrizia Occhineri, Fabio Giannini, Stefania
  Battistini, Claudia Ricci, Michele Benigni, Tea~B. Cau, Daniela Loi, Andrea
  Calvo, Cristina Moglia, Maura Brunetti, Marco Barberis, Gabriella Restagno,
  Federico Casale, Giuseppe Marrali, Giuseppe Fuda, Irene Ossola, Stefania
  Cammarosano, Antonio Canosa, Antonio Ilardi, Umberto Manera, Maurizio
  Grassano, Raffaella Tanel, Fabrizio Pisano, Letizia Mazzini, Sonia Messina,
  Sandra D'Alfonso, Lucia Corrado, Luigi Ferrucci, Matthew~B. Harms, David~B.
  Goldstein, Neil~A. Shneider, Stephen Goutman, Zachary Simmons, Timothy~M.
  Miller, Siddharthan Chandran, Suvankar Pal, George Manousakis, Stanley Appel,
  Ericka Simpson, Leo Wang, Robert~H. Baloh, Summer Gibson, Richard~S. Bedlack,
  David Lacomis, Dhruv Sareen, Alexander Sherman, Lucie Bruijn, Michelle Penny,
  Cristiane de Araujo~Martins Moreno, Sitharthan Kamalakaran, Andrew~S. Allen,
  Braden~E. Boone, Robert Brown, John~P. Carulli, Alessandra Chesi, Wendy~K.
  Chung, Elizabeth~T. Cirulli, Gregory~M. Cooper, Julien Couthouis, Aaron~G.
  Day-Williams, Patrick~A. Dion, Aaron~D. Gitler, Jonathan Glass, Yujun Han,
  Tim Harris, Sebastian~D. Hayes, Angela~L. Jones, Jonathan Keebler, Brian~J.
  Krueger, Brittany~N. Lasseigne, Shawn~E. Levy, Yi~Fan Lu, Tom Maniatis, Diane
  McKenna-Yasek, Richard~M. Myers, Slavé Petrovski, Stefan~M. Pulst, Alya~R.
  Raphael, John Ravits, Zhong Ren, Guy~A. Rouleau, Peter~C. Sapp, Katherine~B.
  Sims, John~F. Staropoli, Lindsay~L. Waite, Quanli Wang, Jack~R. Wimbish,
  Winnie~W. Xin, Hemali Phatnani, Justin Kwan, James~R. Broach, Ximena
  Arcila-Londono, Edward~B. Lee, Vivianna~M. Van~Deerlin, Ernest Fraenkel,
  Lyle~W. Ostrow, Frank Baas, Noah Zaitlen, James~D. Berry, Andrea Malaspina,
  Pietro Fratta, Gregory~A. Cox, Leslie~M. Thompson, Steven Finkbeiner,
  Efthimios Dardiotis, Eran Hornstein, Daniel~J. MacGowan, Terry
  Heiman-Patterson, Molly~G. Hammell, Nikolaos~A. Patsopoulos, Joshua Dubnau,
  Avindra Nath, Rajeeva~Lochan Musunuri, Uday~Shankar Evani, Avinash Abhyankar,
  Michael~C. Zody, Julia Kaye, Stacia Wyman, Alexander LeNail, Leandro Lima,
  Jeffrey~D. Rothstein, Clive~N. Svendsen, Jenny Van~Eyk, Nicholas~J.
  Maragakis, Stephen~J. Kolb, Merit Cudkowicz, Emily Baxi, Michael Benatar,
  J.~Paul Taylor, Gang Wu, Evadnie Rampersaud, Joanne Wuu, Rosa Rademakers,
  Stephan Z{\"{u}}chner, Rebecca Schule, Jacob McCauley, Sumaira Hussain, Anne
  Cooley, Marielle Wallace, Christine Clayman, Richard Barohn, Jeffrey
  Statland, Andrea Swenson, Carlayne Jackson, Jaya Trivedi, Shaida Khan,
  Jonathan Katz, Liberty Jenkins, Ted Burns, Kelly Gwathmey, James Caress,
  Corey McMillan, Lauren Elman, Erik Pioro, Jeannine Heckmann, Yuen So, David
  Walk, Samuel Maiser, Jinghui Zhang, Vincenzo Silani, Cinzia Gellera, Antonia
  Ratti, Franco Taroni, Giuseppe Lauria, Federico Verde, Isabella Fogh, Cinzia
  Tiloca, Giacomo~P. Comi, Gianni Sorar{\`{u}}, Cristina Cereda, Fabiola
  De~Marchi, Stefania Corti, Mauro Ceroni, Gabriele Siciliano, Massimiliano
  Filosto, Maurizio Inghilleri, Silvia Peverelli, Claudia Colombrita, Barbara
  Poletti, Luca Maderna, Roberto Del~Bo, Stella Gagliardi, Giorgia Querin,
  Cinzia Bertolin, Viviana Pensato, Barbara Castellotti, William Camu, Kevin
  Mouzat, Serge Lumbroso, Philippe Corcia, Vincent Meininger, Gérard Besson,
  Emmeline Lagrange, Pierre Clavelou, Nathalie Guy, Philippe Couratier, Patrick
  Vourch, Véronique Danel, Emilien Bernard, Gwendal Lemasson, Hannu
  Laaksovirta, Liisa Myllykangas, Lilja Jansson, Miko Valori, John Ealing,
  Hisham Hamdalla, Sara Rollinson, Stuart Pickering-Brown, Richard~W. Orrell,
  Katie~C. Sidle, John Hardy, Andrew~B. Singleton, Janel~O. Johnson, Sampath
  Arepalli, Meraida Polak, Seneshaw Asress, Safa Al-Sarraj, Andrew King, Claire
  Troakes, Caroline Vance, Jacqueline de~Belleroche, Anneloor~L.M.A. ten
  Asbroek, José~Luis Mu{\~{n}}oz-Blanco, Dena~G. Hernandez, Jinhui Ding,
  J.~Raphael Gibbs, Sonja~W. Scholz, Mary~Kay Floeter, Roy~H. Campbell,
  Francesco Landi, Robert Bowser, Janine Kirby, Roger Pamphlett, Glenn Gerhard,
  Travis~L. Dunckley, Christopher~B. Brady, Neil~W. Kowall, Juan~C. Troncoso,
  Isabelle Le~Ber, Terry~D. Heiman-Patterson, Freya Kamel, Ludo Van Den~Bosch,
  Tim~M. Strom, Thomas Meitinger, Aleksey Shatunov, Kristel van Eijk, Mamede
  de~Carvalho, Maarten Kooyman, Bas Middelkoop, Mattieu Moisse, Russell
  McLaughlin, Michael van Es, Markus Weber, Kevin~B. Boylan, Marka
  Van~Blitterswijk, Karen Morrison, A.~Nazli Basak, Jesús~S. Mora, Vivian
  Drory, Pamela Shaw, Martin~R. Turner, Kevin Talbot, Orla Hardiman, Kelly~L.
  Williams, Jennifer~A. Fifita, Garth~A. Nicholson, Ian~P. Blair, Jesús
  Esteban-P{\'{e}}rez, Alberto Garc{\'{i}}a-Redondo, Ammar Al-Chalabi, Ahmad
  Al~Kheifat, Peter Andersen, Adriano Chio, Jonathan Cooper-Knock, Annelot
  Dekker, Alberto~Garcia Redondo, Marc Gotkine, Winston Hide, Alfredo
  Iacoangeli, Matthew Kiernan, John Landers, Jonathan Mill, Miguel~Mitne Neto,
  Jesus~Mora Pardina, Stephen Newhouse, Susana Pinto, Sara Pulit, Wim
  Robberecht, Chris Shaw, William Sproviero, Gijs Tazelaar, Philip van Damme,
  Leonard van~den Berg, Joke van Vugt, Jan Veldink, Mayana Zatz, Denis~C.
  Bauer, Natalie~A. Twine, Ekaterina Rogaeva, Lorne Zinman, Alexis Brice,
  Eva~L. Feldman, Albert~C. Ludolph, Jochen~H. Weishaupt, John~Q. Trojanowski,
  David~J. Stone, Pentti Tienari, Adriano Chi{\`{o}}, Christopher~E. Shaw, and
  Bryan~J. Traynor.
\newblock {Genome-wide Analyses Identify KIF5A as a Novel ALS Gene}.
\newblock {\em Neuron}, 97(6):1268--1283, 3 2018.

\end{thebibliography}

\end{document}